\documentclass[aps,prapplied,twocolumn,groupedaddress,amsmath,amssymb]{revtex4-2}

\usepackage{graphicx}
\usepackage[dvipsnames]{xcolor}
\usepackage{isotope}

\begin{document}

\title{Generic method to assess transmutation feasibility for nuclear waste treatment and application to irradiated graphite}

\author{Mathieu Rasson}
\email{mathieu.rasson@polytechnique.edu}
\author{Julien Fuchs}
\affiliation{LULI - CNRS, CEA, UPMC Univ Paris 06: Sorbonne Universit\'e, Ecole Polytechnique, Institut Polytechnique de Paris -- F-91128 Palaiseau Cedex, France}

\author{Gr\'egoire Aug\'e}
\email{gauge@onet.fr}
\affiliation{ONET Technologies, Marseille, 13258 Cedex 09, France}

\author{Jean-Louis Qu\'eri}
\affiliation{Department of Physics, Ecole Polytechnique, Institut Polytechnique de Paris -- F-91128 Palaiseau Cedex, France}

\author{G\'erard Laurent}
\affiliation{INSolutions, Chazay d’Azergues, 69380, France}

\date{\today}

\begin{abstract}
\begin{description}
\item[Background] Graphite-moderated nuclear reactors have already produced more than 250,000 tons of irradiated nuclear graphite, or i-graphite, world-wide. The sustainability of this technology relies on the end-of-life management of its moderator, which is activated into a long-lived, low or intermediate-level nuclear waste, by neutron fluxes, during operating time. In particular, carbon-14 is created. Nuclear transmutation, enabled by laser-driven particle acceleration, has been envisioned as a potential novel treatment scheme for long-lived nuclear waste. By triggering controlled nuclear reactions with energetic particles, long-lived radio-nuclides could be transformed into short-lived or stable isotopes. Such a system could treat the carbon-14 nuclei trapped within the i-graphite matrix, an isotope which is difficult to isolate by other means. 
\item[Purpose] This work performs a quantitative preliminary study of this transmutation scheme, in order to assess its feasibility at an industrial scale. The method used can be transposed to assess any transmutation scheme using a beam of particles directly sent on the material to be treated.
\item[Method] First, a nuclear interaction channel which transmutes carbon-14 nuclei without creating new long-lived radio-nuclides is identified. It consists in the choice of a type of particle, among which protons, $\gamma$ photons and neutrons can all be accelerated by laser-matter interaction; and it is completed by the adequate energy at which this particle must be sent on i-graphite. To that end, the nuclear cross-sections of carbon-12, carbon-13 and carbon-14 are reviewed, neglecting other impurities in i-graphite. Then, based on the interaction channel identification, the energy cost of this scheme is estimated.
\item[Results] Protons between $1$ and $5 \, {\rm MeV}$ make it possible to transmute carbon-14 without creating any new long-lived activity.
\item[Conclusions] However, our result show that, even in this favourable reaction channel, the transmutation energy cost is too high for an i-graphite transmutation scheme to be industrially feasible.
\end{description}
\end{abstract}

\maketitle

\section{Introduction}

Fission nuclear energy plays a major role in the limitation of the energy mix carbon print of many countries around the world, such as the United Kingdom, the United States of America, Russia, China and France. Still, it involves another form of environmental toxicity: radioactive and chemical releases, and nuclear waste.

The IAEA nuclear waste management doctrine \cite{trendAIEA} is, by order of preference: recycling, treatment and, only as a last solution, disposal. As reviewed by IAEA GRAPA project \cite{wickham} and by European CARBOWASTE project \cite{carboWaste}, the end-of-life management of irradiated nuclear graphite -- or i-graphite --, a nuclear waste coming from graphite-moderated nuclear reactors (see \cite[][chapter 1]{reuss}) already involving more than 250,000 tons world-wide, is facing difficult issues with carbon-14 treatment. In spite of attempts inspired for example by gas centrifuges \cite{barzashka}, chromatography \cite{cheh}, Pyrolysis / Steam Reforming \cite{steam} or Pressure Swing Adsorption (PSA) \cite{izumi}, no industrial scale scheme for carbon-14 treatment before disposal has been found so far.

Transmutation triggered by laser-driven acceleration has been recently proposed as a possible new treatment scheme for minor actinide transmutation, in a hybrid nuclear reactor \cite{mourou}. The principle is to transform long-lived nuclei into short-lived or stable isotopes, by triggering nuclear reactions with energetic incident particles. By taking advantage of recent progress \cite{danson} in laser chains amplified by the Chirped Pulse Amplification technique (CPA) \cite{mourouCPA}, and in laser-driven particle acceleration \cite{malka}, it is conceivable to design a compact transmutation system which could be transported directly on the site to be treated.

In this context, the goal of this paper is to know whether a laser transmutation scheme can be applied to the particular case of carbon-14 transmutation in i-graphite. First, is developed a generic quantitative assessment method of radio-nuclide transmutation. Then, this method is applied to the particular case of irradiated graphite. The setup studied is as follows. A laser chain amplifies a series of ultra-intense pulses which accelerate particles by laser-matter interaction on an intermediate target. A beam of $\gamma$ photons, protons or neutrons (after proton-neutron conversion, see e.g. \cite{disdier}) is generated in this manner \cite{malka}. Then, the accelerated beam is directly sent on i-graphite in order to transmute carbon-14 nuclei into stable or short-lived isotopes.

Monte-Carlo codes available in the community, such as Geant4, MCNP or Fluka, are designed to simulate complex chains of nuclear reactions. However, some of the radio-nuclides present in i-graphite, or which could be created by transmutation, are not readily described in those codes. Those isotopes have been too scarcely studied. Therefore, because of a lack of cross section data, those codes are not directly suited to describe irradiated graphite transmutation. In order to address this issue, a dedicated simulation tool has been developed to incorporate the state-of-the-art nuclear cross section data available for those radio-nuclides, i.e. the Evaluated Nuclear Data Files on the IAEA website \cite{endf}, including TALYS simulations \cite{tendl}.

In order to focus on the issue of carbon-14 treatment within the graphite matrix, a mix of carbon isotopes only is considered. No other radio-nuclide coming from i-graphite impurities is treated. This mix consists in natural proportions of carbon-12 and carbon-13, $98.93 \, \%$ and $1.07 \, \%$ respectively, completed by a fraction of carbon-14 following the Poncet \cite{poncet} inventory, $660 \, {\rm ppb}$.

In this nuclear mix, the nuclear reactions triggered by $\gamma$ photons, protons and neutrons are to be studied in the following sections, in order to find the particle and the energy which can efficiently transmute carbon-14 without creating new radio-nuclides. This combination of parameters defines a reaction channel. Then, the energy cost of this scheme is to be estimated in order to establish whether this solution is feasible at an industrial scale.

\section{Method and Modeling}

\subsection{Population evolution}

For clarity, two distinctions must be introduced. First, a nuclear reaction is said \emph{primary} when it involves an incident particle and a nuclei of the target. Such a reaction can release energetic particles in the target if the incident particle has an excess of kinetic energy compared to the energy threshold of the reaction. Therefore, the energetic products can themselves trigger other nuclear reactions. Those reactions are said \emph{secondary}. Second, the goal of this analysis is to predict the evolution of nuclear populations inside a target under a flux of incident particles. As soon as the first nuclear reactions occur, new isotopes appear in the target which can be themselves transformed by incident particles. Therefore, to predict the evolution of the composition of the target, all reactions between incident particles and initial nuclei in the target must be known, but reactions with nuclei created by anterior reactions must be known as well. Such a treatment of the target is said \emph{dynamic}. Still, in order to have an idea of the parameters required to initiate transmutations and of the cost of those first reactions, a target containing only the initial nuclei can be considered. Such a target is said \emph{static}. The advantage of this simplified framework is to present an easily readable evolution from the initial conditions, without any complex coupling.

In all this study, as will be justified later, only primary reactions are accounted for, and the i-graphite target is considered static.

The probability of occurrence of a nuclear transmutation reaction when a particle $A$ is sent, with a certain kinetic energy, on a nucleus $B$, at rest, is described by the microscopic cross section of this reaction. The number of reactions per time and volume units inside the target is given by $j_A \sigma(\varepsilon) n_B$, where $j_A$ is the norm of the flux of particles $A$ (number of particles per surface and time units) going through the target with a kinetic energy $\varepsilon$, $n_B$ is the density of nuclei $B$ inside the target and $\sigma(\varepsilon)$ the microscopic nuclear interaction cross section at $\varepsilon$ (see e.g. \cite[][page 54]{lamarsh}).

From this local expression, two assumptions are necessary to draw a global form. First, only a region of volume $V$ in which the flux $j_A$ is homogeneous is considered. Then, the evolution of this region under the flux of incident particles is assumed isochoric. At all times, inside this region, the population of B, $N_B$, is linked to its density by $N_B = n_B V$.

Therefore, the population $N_B$ is governed by the following linear first order differential equation:
\begin{equation}
\dfrac{{\rm d} N_B}{{\rm d}t} = - j_A \sigma(\varepsilon) \, N_B
\label{eq:population_monoE}
\end{equation}

In a static target framework, this decay law is valid if the target contains only $B$ and as well if it contains a mix of nuclides. Such an equation can be written for each type of nucleus inside the target.

So far, only a mono-energetic flux of $A$ has been considered. In order to account for realistic beam spectra in this linear model, the transmutation contribution of each spectral component has to be summed. The total flux $j_A$ can be written $j_A = \int_\mathcal{S} \rho_A(\varepsilon) {\rm d}\varepsilon$, where $\rho_A$ is a flux spectral density (number of particles per units time, surface and energy), and $\mathcal{S}$ is the spectrum of the flux, supporting the spectral distribution $\rho_A$. With those notations, the mono-energetic population evolution Eq.~\ref{eq:population_monoE} can be rewritten in the following way:
$$
\dfrac{{\rm d} N_B}{{\rm d}t} = - \left( \int_\mathcal{S} \rho_A(\varepsilon) \sigma(\varepsilon) {\rm d}\varepsilon \right) \, N_B
$$

\subsection{Decay parameters}

The decay of population $B$ can effectively be described by a half-transmutation time, that is to say the time needed by the flux to divide the targeted population by two.

As this description is only valid for a particle flux constant with time, it is necessary to find a more general formulation adapted to pulsed deposition of particles. Such a formulation can be obtained simply by a variable change in Eq.~\ref{eq:population_monoE}. Is introduced the number of particles sent on the target per surface unit between date $0$ and date $t$, $t \mapsto J(t)$. Its time differential is the number of particles sent per time and surface units, i.e. $j_A$. When inserting a $J$ dependence in Eq.~\ref{eq:population_monoE}, the right member is divided by $j_A$, which absorbs every explicit time dependence. When the source has a non trivial spectrum, the corresponding equation is
$$
\dfrac{{\rm d} N_B}{{\rm d}J} = - \left( \int_\mathcal{S} \dfrac{\rho_A(t, \varepsilon)}{j_A(t)} \sigma(\varepsilon) {\rm d}\varepsilon \right) \, N_B
$$
where it must be stressed that the fraction $\rho_A(t, \varepsilon)/j_A(t)$ does not depend on time. The flux spectrum only varies in global amplitude with time, which means that the spectrum distribution normalised by the total flux is constant.

Thus, the decay of a population has no explicit reference to time. The characteristic decay time is in this case replaced by a characteristic number of particles to send per surface unit in order to divide the population $B$ by two. Its expressions with a non-trivial spectrum is as follows:
$$
J_{1/2}^{-1} = \dfrac{1}{\ln(2)} \, \int_\mathcal{S} \dfrac{\rho_A(\varepsilon)}{j_A} \sigma(\varepsilon) {\rm d}\varepsilon
$$

Finally, by multiplying the number of particles sent to perform a half-transmutation in the considered region by the energy of the mono-energetic beam or by the mean energy of its spectrum, a half-transmutation fluence is obtained. This quantity measures the energy per surface unit which has to be given to the particle beam in order to perform a half transmutation:
$$
S_{1/2} = \mathcal{E}_0 \, J_{1/2}
$$
where $\mathcal{E}_0$ is alternatively the energy of a mono-energetic beam or the average energy of a beam with a spectral distribution, $\mathcal{E}_0 = \int_\mathcal{S} \varepsilon \rho_A(\varepsilon)/j_A {\rm d}\varepsilon$.

\subsection{Flux penetration}

The derivation presented in the previous sections relies on a strong assumption regarding the incident particle flux: it has to be homogeneous in the region considered. Actually, both the particle flux and its energy decrease while propagating inside the target. The energy of a neutral particle, a neutron or a $\gamma$ photon in this case, is dissipated as a decreasing exponential, and that of a charged particle, a proton here, is absorbed according to Coulomb collisions ending on a Bragg peak \cite{bragg1}, \cite{bragg2}. Therefore, the number of transmutations predicted by the previous expressions is only valid on the penetration depth of the incident flux. The values used in this paper come from Geant4 \cite{geant4} simulations cross-checked with the code ESTHER \cite{esther} for the proton channel. In Geant4, High Precision libraries based on \texttt{QGSP\_BIC\_HP} are used for hadrons with standard electromagnetic libraries. The graphite detector is composed of a natural carbon mix, as carbon-14 is negligible for beam energy dissipation, with a $1.7 \, {\rm g/cm^3}$ volumetric mass, i.e. typical value for a nuclear graphite sample \cite{tecdocAIEA}.

This study has mainly been done for protons, as suggested by reaction channels which will be discussed later on. For simplicity, the energy deposition is assumed infinitely localised, and the incident flux is homogeneous over the depth preceding its total absorption. As the position of this Bragg peak depends strongly on the energy of the incident particle, average penetration depths are calculated for a beam with a realistic spectrum, as follows:
$$
\text{penetration depth} = \int_\mathcal{S} \delta(\varepsilon) \dfrac{\rho_A(\varepsilon)}{j_A} \, {\rm d}\varepsilon
$$
where $\delta(\varepsilon)$ is the penetration depth of the particle $A$ at energy $\varepsilon$, $\rho_A$ is the flux spectral distribution of the beam considered on the spectrum $\mathcal{S}$ with total flux $j_A$.

With this penetration depth known, it is possible to convert the half-transmutation fluence into a transmutation energy cost by mass unit:
$$
\text{mass cost} = \dfrac{\text{fluence}}{\text{density} \times \text{penetration depth}}
\label{eq:cost}
$$

The cost determined here corresponds to an energy in the beam of particles, and not to wall plug energy. It does not take into account the conversion efficiency of the method used to generate energetic particles. Working with beam energy has the advantage of being independent of the particular technological solution used to accelerate particles. This cost is only determined by the nuclear properties of the isotopes to be treated. Integrating the wall plug-particles efficiency would obviously make the challenge, in terms of practical feasibility, greater.

\subsection{Data and Channel identification}

In order to obtain a readable comparison of cross sections in i-graphite, as mentioned before, a static target is considered. All reactions between carbon-12, carbon-13, carbon-14 and incident $\gamma$ photons, neutrons and protons are accounted for between $0$ and $20 \, {\rm MeV}$ with an extension up to $30 \, {\rm MeV}$ for neutrons.

Those cross sections are taken from Evaluated Nuclear Data Files on the IAEA website \cite{endf} and from the TALYS library \cite{tendl} when no evaluated data was found.

The reactions listed in this work are grouped into two categories. A transmutation reaction is a reaction which involves carbon-14 and which transforms it into another nucleus. A useless reaction is a reaction which involves carbon-12 or carbon-13. Then, from those two classes, are qualified harmful all reactions which create a new radioisotope whose half-life is greater than a year, tritium (12.3 years) or beryllium-10 (1.4 million years) in this case.

\subsection{Proton spectrum}

We will focus in this subsection on the specific case of a proton beam as the input beam. The particular interest of protons will be justified below.

In order to derive more accurate efficiency and energy cost estimations about the proton channel, it is necessary to consider a beam with a realistic spectrum, and not only mono-energetic incident particles. Such spectrum is well described by an exponential fit \cite{macchi}:
$$
\rho_A(\varepsilon) = \dfrac{j_A}{k_{\rm B} T} \, \exp\left( - \dfrac{\varepsilon}{k_{\rm B} T} \right)
$$
where $j_A$ is the total flux (number of particles per time and surface units), $k_{\rm B}$ the Boltzmann constant, $T$ the temperature of the beam \cite{snavely} and $\varepsilon$ the energy of the spectral component considered.

Note that this spectrum has necessarily a high energy cut-off. As shown by Fuchs \cite{fuchs}, such cut-off can be chosen independently of the temperature, thanks to acceleration design.

In this paper, which focuses on nuclear channels, the proton energy cut-off is set arbitrarily to match cross-section data. In this case, there is only one free parameter which defines the proton spectrum: the proton temperature.

The use of a laser-driven accelerated method is only a particular case of application of the generic assessment method presented. In order to account for another source of energetic particles, the spectrum generated has to be adapted.

\subsection{Generic Method Summary}

To summarize, the generic transmutation assessment method presented in this work consists in the following steps.

\begin{enumerate}
\item List the nuclear reactions offered by the incident particles considered, with their respective cross sections, and categorise them between transmutation, useless and harmful reactions;
\item Identify, if possible, a favourable reaction channel in which transmutation is possible without any additional long-lived activity;
\item Estimate the penetration depth of the favourable incident particle, at the corresponding energy, in the target to be treated;
\item Compute the transmutation energy cost per mass unit, in beam energy.
\end{enumerate}

In order to refine the cost obtained, the actual spectrum generated by the source of energetic particles envisioned can be used. Besides, it can highlight potential harmful reactions triggered by spectral distribution tails.

\section{Results}

As irradiated nuclear graphite is a mixture of carbon isotopes with strong differences in concentration between those, macroscopic cross sections are more relevant to estimate the relative importance of each reaction. With the notations used before, it can be written $\Sigma = n_B \sigma$.

\begin{figure}
    \includegraphics[width=8.6cm]{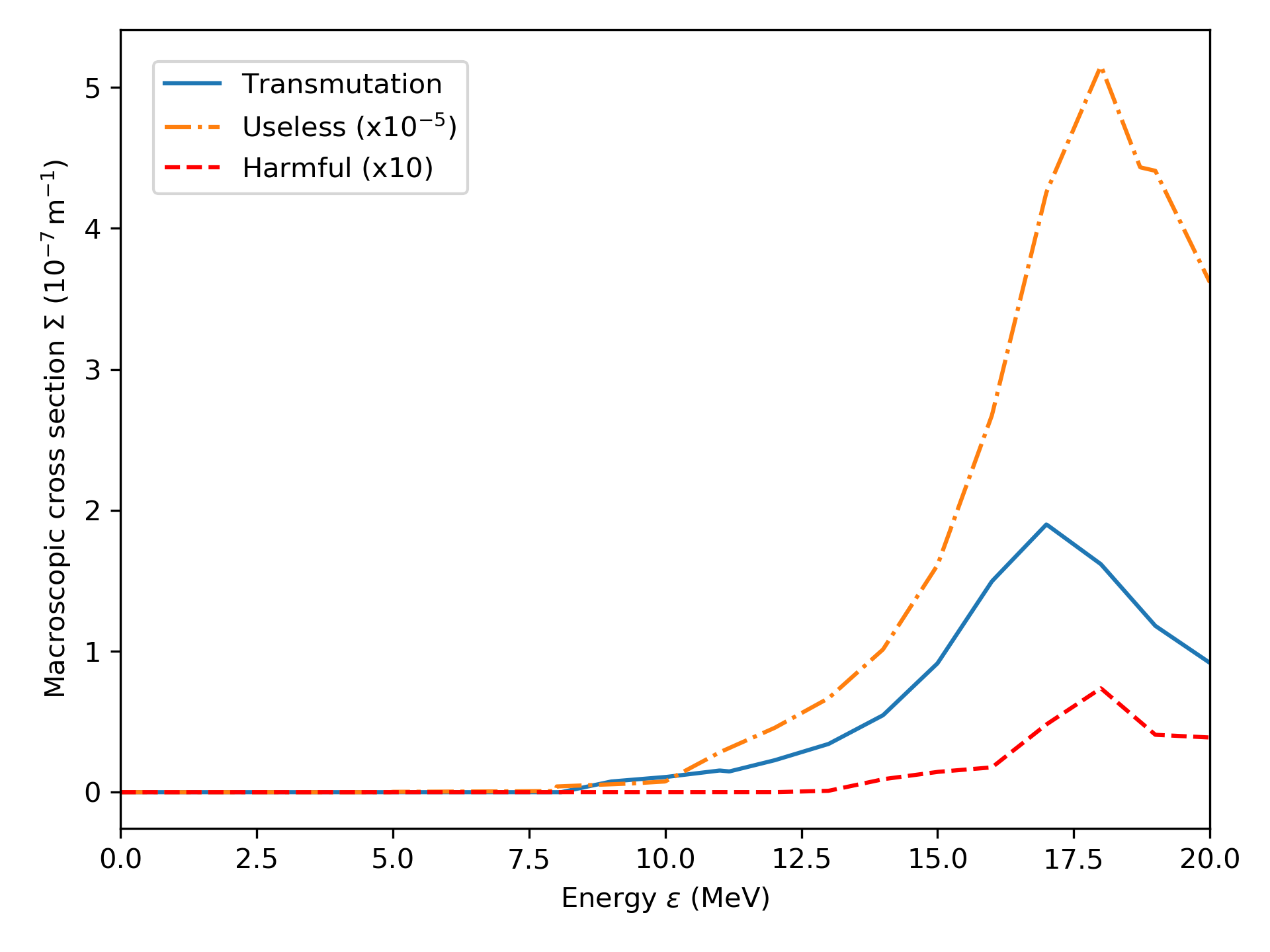}
    \caption{Macroscopic cross sections in irradiated nuclear graphite under incident $\gamma$ photon mono-energetic flux, for various beam energies.}
    \label{fig:cross_gamma}
\end{figure}

\begin{figure}
    \includegraphics[width=8.6cm]{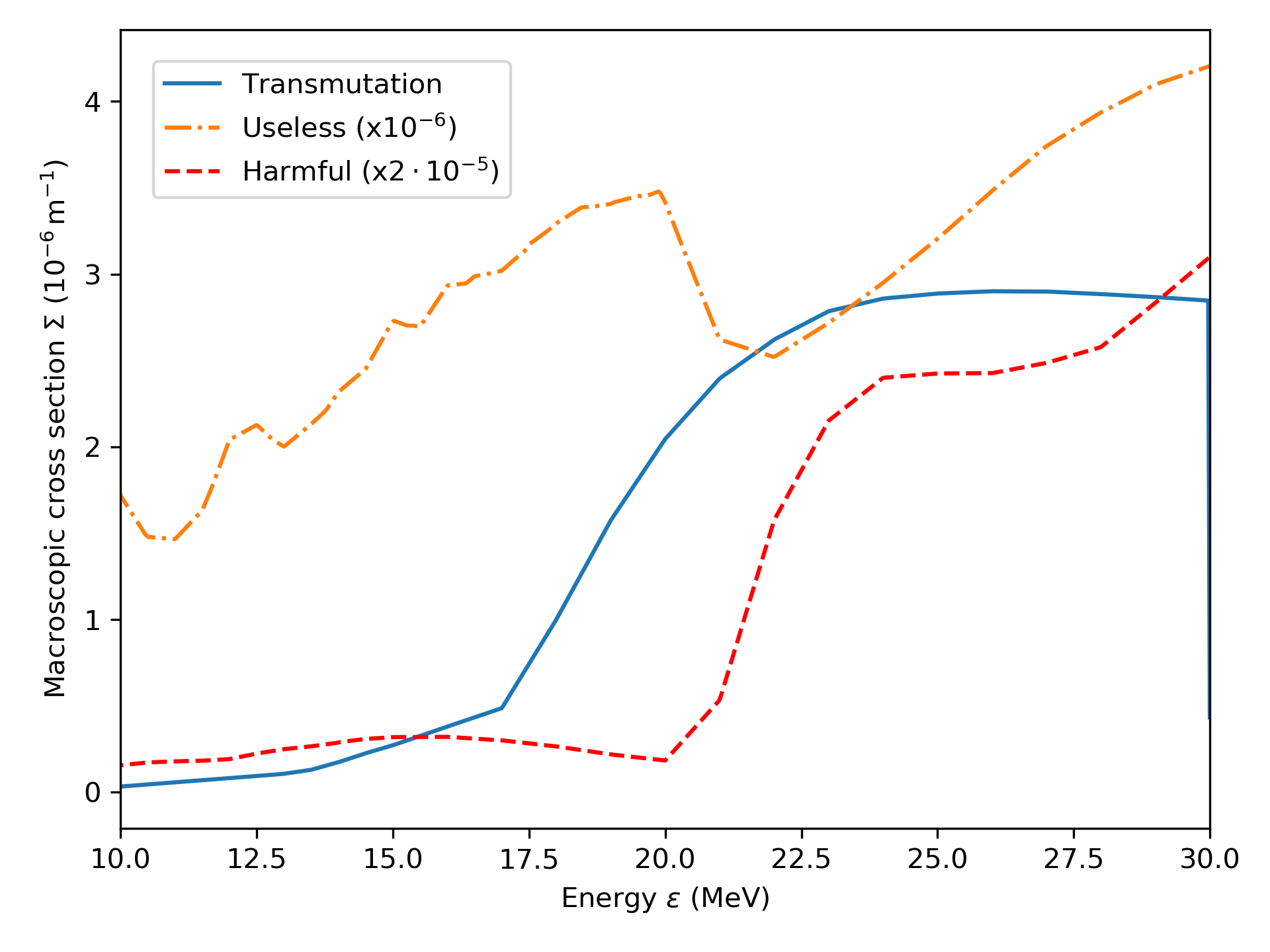}
    \caption{Macroscopic cross sections in irradiated nuclear graphite under incident neutron mono-energetic flux, for various beam energies. Data for $\isotope[12]{C} (n, 3\alpha)$ have been extrapolated between $20$ and $30 \, {\rm MeV}$, and classified as useless cross section.}
    \label{fig:cross_neutron}
\end{figure}

\begin{figure}
    \includegraphics[width=8.6cm]{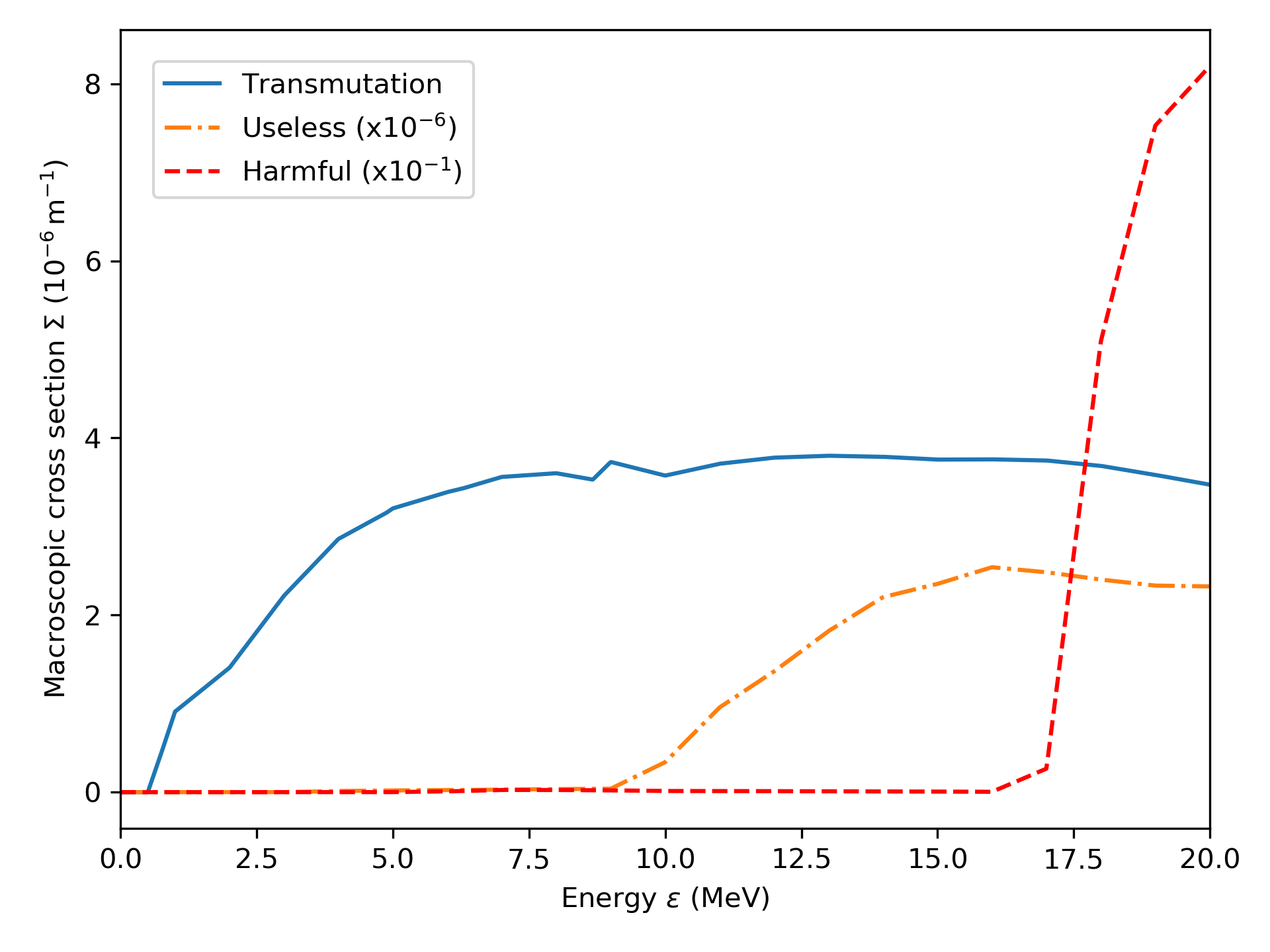}
    \caption{Macroscopic cross sections in irradiated nuclear graphite under incident proton mono-energetic flux, for various beam energies.}
    \label{fig:cross_proton}
\end{figure}

In Figures \ref{fig:cross_gamma}, \ref{fig:cross_neutron} and \ref{fig:cross_proton}, the macroscopic cross sections offered by $\gamma$ photons, neutrons and protons in i-graphite are displayed in their respective categories: transmutation, useless and harmful. For neutrons, because of a lack of data on the $\isotope[12]{C} (n, 3\alpha)$ reaction between $20$ and $30 \, {\rm MeV}$, data between $0$ and $20 \, {\rm MeV}$ have been extrapolated following a decrease in $1/\sqrt{\varepsilon}$ up to $30 \, {\rm MeV}$. This dependency is chosen to be coherent with the general decrease tendency of inelastic neutron cross sections in $1/v$, where $v$ is the velocity of the incident neutron (see e.g. \cite[][page 54]{reuss}). This rough extrapolation concerns a useless but not harmful reaction, so it does not have a real impact on the channel choice.

In contrast to the to other particles, mono-energetic protons below $4 \,{\rm MeV}$ can transmute carbon-14 nuclei without creating any new radio-nuclide, and harmful cross sections remain limited for protons below $17 \,{\rm MeV}$. Both neutrons and $\gamma$ photons lead to the creation of beryllium-7 nuclei while transmuting carbon-14 nuclei. In Figure \ref{fig:cross_neutron}, in particular, the compression of the harmful cross section axis hides the large quantity of new radio-nuclides created in the energy window explored.

Overall, it appears that mono-energetic protons below $4 \,{\rm MeV}$ offer the only favourable solution, avoiding any additional long-lived activity.

Still, it must be noted that protons suffer from a very low penetration depth around $5 \, {\rm MeV}$, of the order of $100 \, {\rm \mu m}$, when neutrons of $20 \, {\rm MeV}$ can propagate on a centimetre scale. Nevertheless, this difference can easily be compensated by the production efficiency of those particles. As many neutron production schemes use accelerated protons to create neutrons, neutrons introduce an additional conversion step. The best performance so far in terms of yield \cite{kleinschmidt} has reported more than $5 \cdot 10^{10}$ total neutrons, generated from a proton beam having a total yield around $10^{13}$ \cite{wagner}, so a proton-to-neutron yield of the order of $1/200$.

Now, to estimate the overall cost of transmutation with protons -- the most favourable channel --, the mono-energetic cross section data presented before have been convoluted with a typical exponential proton spectrum, such that the temperature of the proton beam becomes the new energy incident parameter. This exponential spectrum is truncated at $20 \, {\rm MeV}$ to fit cross section data.

\begin{figure}
    \includegraphics[width=8.6cm]{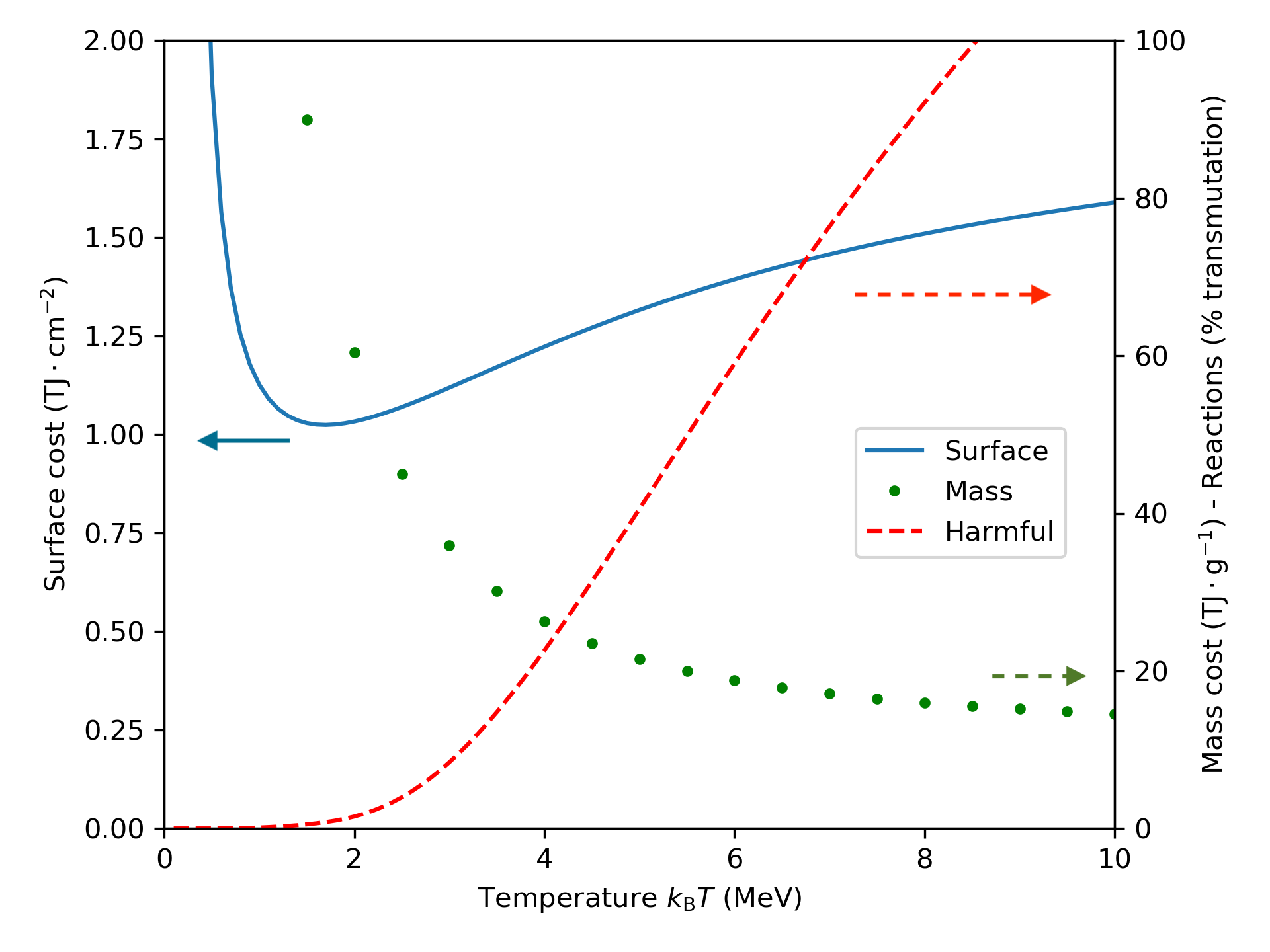}
    \caption{Irradiated nuclear graphite transmutation mass (right axis) and surface (left axis) beam energy cost, under incident proton flux having exponential spectrum of various temperatures. As context, the ratio harmful over transmutation macroscopic cross sections is plotted for those temperatures (right axis).}
    \label{fig:cout}
\end{figure}

The results of those calculations are displayed in Figure \ref{fig:cout}. As a function of the proton temperature, the continuous blue curve represents the transmutation surface cost, or half-transmutation fluence, on the left axis. With the right axis, is drawn as green dots the half-transmutation mass cost. And, with the right axis again, used as \% and not $\rm TJ/g$, is represented as a red dashed line the proportion of harmful reactions over transmutation reactions.

It can be seen that the proportion of harmful reactions increases rapidly with the beam temperature. This is due to some protons within the distribution having individual kinetic energy above $17.5 \, {\rm MeV}$, and those being able to trigger harmful reactions.

Lastly, Figure \ref{fig:cout} gives an order of magnitude of the half-transmutation mass cost with the proton channel, in beam energy: $20$ to $40 \, {\rm TJ/g}$, i.e. $5$ to $11 \, {\rm GWh/g}$. For illustration, let's consider the French nuclear reactor Bugey 1, which belongs to the Uranium Naturel Graphite Gaz technology, moderated by graphite. During its 22 years of exploitation, at $540 \, {\rm MW}$ of electricity generation \cite{bugey}, it produced $3.7 \cdot 10^5 \, {\rm TJ}$ of electrical energy. If this entire energy production was invested in proton pulses, it could only perform a half transmutation on $10 \, {\rm kg}$ of i-graphite, when more than 2,000 tons must be treated \cite{poncet}.

It must be highlighted that this cost is expressed in beam energy, only constrained by the nuclear properties of carbon-14. All the conversion inefficiencies which would be introduced by the technological solution used to accelerate particles would have to be added to this cost.

Therefore, because of the nuclear properties of i-graphite which require the use of protons as energetic incident particles, because of the low cross sections involved and because of the shallow penetration depth of protons, transmuting carbon-14 nuclei in i-graphite is not energetically feasible.

\section{Discussion}

The results presented in the last section, and in particular the cost estimation, rely on the hypotheses of the linear model introduced. First, secondary reactions are neglected, which reduces the number of reactions taken into account. The role played by those nuclear reactions induced by products has been estimated independently thanks to Geant4 simulations -- with the same libraries as before --, and are of the order of $1 \,\%$ of the primary reactions. Taking into account secondary reactions cannot compensate for the several orders of magnitude gain in efficiency required for i-graphite transmutation to be feasible. Nevertheless, if, for a certain radio-nuclide to be treated, this linear model gives encouraging results, it would be interesting to consider those reactions, in order to refine the assessment of the transmutation cost.

Second, the target is considered static in this estimation, which means that only reactions with carbon-12, carbon-13 and carbon-14 are accounted for. In this case, all carbon-14 transmutation reactions are taken into account, so there is no underestimation of transmutation efficiency on this regard. Solely reactions which could create new carbon-14 nuclei are potentially neglected. It appears with an annex dynamic treatment, with the same cross section sources as described above, that there is only one reaction in this category in the energy range reviewed, $\isotope[15]{N} (p, 2p) \isotope[14]{C}$, and that the low proportion of nitrogen-15 created, five orders of magnitude below typical carbon-14 concentrations, prevents it from playing an important role.

Third, this model needs an isochoric assumption. This hypothesis is hardly justifiable when the population of the target begins to be significantly modified. However, it is likely that energy and particle deposition inside the target would rather increase its volume and dilute the nuclei to be treated. This effect would tend to play against transmutation, increasing the waste volume to be treated and therefore increasing the total transmutation cost.

Finally, cross section data is possibly the major shortcoming of this work. The results given by the assessment method described here, by focusing on the nuclear physics constraints of transmutation, crucially rely on the cross sections used as input. As mentioned before, isotopes such as carbon-14 or beryllium-7 have scarcely been studied experimentally, and the data used in this article come from numerical simulations which have not been compared to experiments. However, the library used has been tabulated from the TALYS code, which is the reference in this field.

More precisely, the work of Sanders \cite{sanders} suggests the existence of resonances in interesting transmutation reactions and in the energy range investigated here, which are not accounted for by TALYS data. Still, for the reaction $\isotope[14]{C} (p,n)$ studied by Sanders, the highest resonance peak reaches a cross section of $650 \, {\rm mb}$, at $2.08 \, {\rm MeV}$, when TALYS predicts a cross section going from $160 \, {\rm mb}$ to $260 \, {\rm mb}$ between $1 \, {\rm MeV}$ and $3 \, {\rm MeV}$. Consequently, even with a mono-energetic incident beam perfectly aligned with the resonance, only a factor 3 to 4 could be gained in transmutation efficiency. This gain remains far from the orders of magnitude necessary to make industrial graphite transmutation feasible. However, in more litigious cases, the effect of resonances would not be negligible and more precise cross section data would be required.

It must be noted that, on Figure \ref{fig:cout}, through the spectrum of the proton beam generated, the particular choice of laser-driven acceleration has an impact on the transmutation cost calculated. However, because of the proton transmutation cross section profile -- slowly varying with a low energy threshold -- the transmutation surface costs with mono-energetic protons and with protons having an exponential spectrum do not differ greatly. The minimum of this cost with an exponential spectrum is around $1 \, {\rm TJ \cdot cm^{-2}}$ while it is around $0.75 \, {\rm TJ \cdot cm^{-2}}$ with mono-energetic particles. Therefore, the particular choice of laser-driven acceleration to generate protons does not alter the general result of this study: nuclear physics imposes a very high transmutation energy cost for carbon-14 transmutation.

\section{Conclusion}

The major issue of nuclear graphite life cycle management resides in the end of life treatment of the large amount of i-graphite created by graphite-moderated nuclear reactors. In particular, the long-lived carbon-14 nuclei trapped within the graphite matrix and mixed with other carbon isotopes must be dealt with, if the volume of i-graphite is to be reduced.

This work shows that direct i-graphite transmutation with protons of an energy lower than $4 \, {\rm MeV}$ displays the most interesting nuclear properties. However, even in this favourable reaction channel, the transmutation energy cost is too high for a reasonable i-graphite industrial transmutation scheme to be successful.

The method used here to assess the feasibility of carbon-14 transmutation can be transposed to any transmutation scheme sending directly a beam of particles on the material to be treated. For instance, this beam-target scheme, to be distinguished from transmutation in a reactor (see e.g. the MYRRHA project \cite{abderrahim2020}), has been envisioned to transmute caesium-137 \cite[][chapter 1]{nakajima} and iodine-129 \cite{magill}, without quantitative assessments. A systematic identification of transmutation channels and estimation of the corresponding treatment cost could give insightful orientations towards the opportunities of long-lived nuclear waste transmutation, by discarding schemes prohibitively expensive in energy.

\bibliography{biblio}

\end{document}